**Title**

# Hydrologic Cycle Weakening in Hothouse Climates

**Authors**

Jiachen Liu[1], Jun Yang[1]*, Feng Ding[1], Gang Chen[2], Yongyun Hu[1]

**Affiliations**

[1]Laboratory for Climate and Ocean-Atmosphere Studies, Department of Atmospheric and Oceanic Sciences, School of Physics, Peking University; Beijing 100871, China.

[2]Department of Atmospheric and Oceanic Sciences, University of California-Los Angeles; Los Angeles, CA 90095, USA.

*Corresponding author. Email: junyang@pku.edu.cn

**Abstract**

The hydrologic cycle has wide impacts on the ocean salinity and circulation, carbon and nitrogen cycles, and the ecosystem. Under anthropogenic global warming, previous studies showed that the intensification of the hydrologic cycle is a robust feature. Whether this trend persists in hothouse climates, however, is unknown. Here we show in climate models that mean precipitation first increases with rising surface temperature, but the precipitation trend reverses when the surface is hotter than ~320-330 K. This non-monotonic phenomenon is robust to the cause of warming, convection scheme, ocean dynamics, atmospheric mass, planetary rotation, gravity, and stellar spectrum. The weakening occurs because of the existence of an upper limitation of outgoing longwave emission and the continuously increasing shortwave absorption by $H_2O$, and is consistent with atmospheric dynamics featuring the strong increase of atmospheric stratification and dramatic reduction of convective mass flux. These results have wide implications for the climate evolutions of Earth, Venus, and potentially habitable exoplanets.

**Teaser**

Climate models predict that mean precipitation is a non-monotonic function of temperature and weakens under hothouse climates.

**Introduction**

In the study of climate change, one of the key findings is that global-mean precipitation ($P_m$) should increase with surface temperature ($T_s$), which is supported by numerous climate simulations, theoretical derivations, and observations (e.g., (*1–4*)). This conclusion was obtained from studies of climates that are relatively close to modern Earth, typically within $\pm 10$ K. Whether the conclusion can be applied to much hotter climates is unknown. In our study, we reveal the presence of a critical transition point in the relationship between $P_m$ and $T_s$. Below this threshold, $P_m$ exhibits an increasing trend with $T_s$ ($dP_m/dT_s>0$), whereas above it, the trend reverses, featuring a reduction in $P_m$ with $T_s$ ($dP_m/dT_s<0$). Remarkably, for Earth's atmosphere and orbit, this transition from intensification to weakening occurs at around 320-330 K. This suggests that the previously discovered conclusion cannot be generalized to hothouse climates.

Over a long-term average, $P_m$ is determined by surface or atmospheric energetic constraint (e.g., (*5–8*)). Quantitatively, outgoing longwave radiation to space ($OLR$) is balanced by net longwave radiation from the surface ($NLW^s$), atmospheric shortwave absorption ($ASW^a$), sensible heat flux from the surface ($SH^s$), and condensation heating from precipitation ($L\rho_w P_m$), written as:

$$L\rho_w P_m = OLR - NLW^s - ASW^a - SH^s, \qquad (1)$$

where $L$ is the latent heat of vaporization, $\rho_w$ is the liquid water density, and ($OLR - NLW^s$) is atmospheric net longwave emission. The dominant terms are $L\rho_w P_m$, $OLR$, $NLW^s$, and $ASW^a$, while the term $SH^s$ is smaller. For the modern Earth, the four terms on the right-hand side of Eq. (1) are respectively 240±3, 52±9, 75±10, and 24±7 W m$^{-2}$ (*9*), and the estimated $P_m$ is ~3.09±0.35 mm day$^{-1}$, close to reanalysis data. Previous studies showed that $P_m$ increases with $T_s$ and the increasing rate is around 2-3%/K (e.g., (*3,4*)). The main reason for the intensification of precipitation is the increase of $OLR - NLW^s$ with $T_s$ (*10*), which is dominated by the decrease of $NLW^s$ (*11*).

At low temperatures, $OLR$ is an approximately linearly increasing function of $T_s$ (*12*). At high temperatures, however, $OLR$ asymptotes to a limiting, maximum value ($dOLR/dT_s = 0$) when the atmosphere becomes optically thick in all infrared wavelengths and only thermal radiation from the upper troposphere can emit to space (*13, 14*). For instance, the limiting $OLR$ is ~282 W m$^{-2}$ for a pure water vapor atmosphere (*15*). However, $ASW^a$ should continue to increase with $T_s$ because the atmospheric absorption in shortwave wavelengths has not reached saturation. The shortwave absorption predominately depends on water vapor concentration, which increases exponentially with temperature following the Clausius-Clapeyron relation. These different trends of longwave cooling and shortwave absorption when the surface is warmer than 320 K were depicted in Jeevanjee & Romps (*10*). Inspired by the insightful work of Jeevanjee & Romps (*10*), we speculate that $P_m$ would decrease with $T_s$ *($dP_m/dT_s < 0$)* when the increasing rate of $ASW^a$ exceeds that of $OLR$ in hothouse climates. Below, we test this hypothesis through a hierarchy of climate model experiments: global three-dimensional (3D) climate simulations using three different atmospheric general circulation models (GCMs), small-domain 3D radiative-convective simulations using two cloud-resolving models (CRMs), and one-dimensional (1D) radiation calculations using a radiative transfer model (RTM), as described in table S1 and Materials and Methods.

## Results

### The trend of mean precipitation

$P_m$ initially increases with $T_s$ and then decreases (Fig. 1). The transition from an increasing to a decreasing trend occurs within a range of ~320 to 330 K. The increasing trend of $P_m$ below the transition temperature is consistent with previous studies (*1-4, 16*), whereas the decreasing trend beyond the transition temperature (*10*) is a less-explored phenomenon emphasized in this study. The non-monotonic relationship between $P_m$ and $T_s$ is found in both global climate simulations with parameterized convection and clouds (Fig. 1, A, B, and C) as well as small-domain cloud-resolving simulations with explicit convection and clouds (Fig. 1D). The trend is independent of the cause of warming due to increasing $CO_2$ (Fig. 1, B and C), raising solar constant (Fig. 1A), or specifying $T_s$ (Fig. 1D). However, the cause of warming does influence the maximum $P_m$ (ranging 3.5 and 5.5 mm day$^{-1}$) and the transition temperature. The transition temperature is lower as $CO_2$ concentration is higher (Fig. 1B), as $CO_2$ acts to reduce $OLR$, consequently lowering $P_m$ for a given $T_s$ (Eq. (1)). In addition, the non-monotonic trend is insensitive to land-sea configuration, as the experiments in Fig. 1A and B were conducted with modern Earth's land-sea configuration or reconstructed Earth's paleography ~630 million years ago while the experiments in Fig. 1C and D were coupled with surface ocean everywhere. These results suggest that the weakening of $P_m$ under hothouse climates is robust.

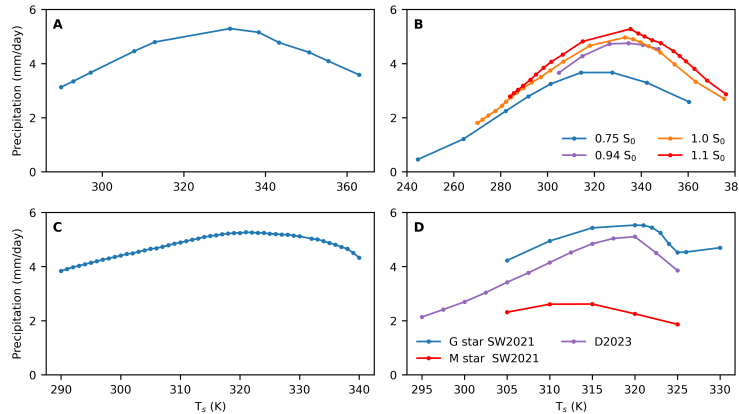

**Fig. 1. Mean surface precipitation as a function of mean surface temperature ($T_s$) simulated by three global climate models (A, B, C) and by two cloud-resolving models (D).** (**A**) ExoCAM experiments with increasing solar constant (*18*). (**B**) ExoCAM (*48*) and CAM3 experiments with increasing $CO_2$ concentration under 0.75 (blue), 0.94 (purple), 1.0 (orange), and 1.1 (red) times modern solar constant (S$_0$). For a given $T_s$, the input $CO_2$ concentration is higher for a lower solar constant. (**C**) ExoCAM experiments with increasing $CO_2$ concentration for an aqua-planet with modern solar constant (*49*). (**D**) Small-domain fixed-SST (sea surface temperature) cloud-resolving experiments using DAM (*22*) and SAM (*45*). Blue and red lines show results with the solar spectrum (blue) and an M star spectrum (red) in ref. (*22*) and the purple line depicts results from ref. (*45*). Mean precipitation is a non-monotonic function of $T_s$ and weakens under hothouse climates.

Based on the surface energy budget, previous studies of refs. *6-8 & 17* suggested that global-mean precipitation is a monotonic function of surface temperature and asymptotically approaches a maximum value at ~320 K. Here we show that the global-mean precipitation is a non-monotonic function and decreases with increasing surface temperature in hothouse climates. One reason for the difference is that previous studies did not explore climates warmer than 320 K. Another reason is that shortwave absorption was not included in the idealized gray radiation GCM simulations of O'Gorman & Schneider (*7*), whereas, in our simulations, both shortwave and longwave radiation calculations are included. Based on analytical argument and simulation data from a small-domain cloud-resolving model, Jeevanjee & Romps (*10*) suggested that atmospheric net radiative cooling rate should decrease with surface temperature when the surface is warmer than 320 K. Our finding confirms their prediction, explicitly demonstrating the decreasing trend. Moreover, we extend the phenomenon to GCMs and apply it to other planets (see below "Applying to other planets"), further proving the robustness of this conclusion.

How to understand the non-monotonic trend of $P_m$ as a function of $T_s$? Below, we answer this question through two distinct approaches: energetic constraint and dynamical constraint, as shown in the schematic diagram presented in fig. S1.

**Mechanisms: energetic constraint and dynamical constraint**

In the context of energetic constraint, energy conservation states that the amount of atmospheric latent heat release (related to $P_m$) is determined by the ability of atmospheric energy loss (Eq. (1)). Below the transition temperature, $P_m$ increases with $T_s$ mainly because $OLR - NLW^s$ increases with $T_s$, but the increasing rate becomes smaller as $T_s$ is close to the transition temperature. Above the transition temperature, $P_m$ decreases with $T_s$ ($dP_m/dT_s < 0$), because $OLR$ asymptotes to a constant value when the atmospheric optical thickness becomes large enough (Fig. 2B) and meanwhile atmospheric shortwave absorption ($ASW^a$) keeps increasing with $T_s$ (Fig. 2C). The strong increasing of $ASW^a$ overcomes the change of $OLR - NLW^s$ and other factors. The different trends between longwave and shortwave are because atmospheric cross-sections of water vapor in visible and near-infrared wavelengths are less than those in thermal infrared wavelengths (see Figure S3 in Goldblatt et al. (*15*)). As seen in fig. S2B, in moderate climates, atmospheric absorption in visible and near-infrared wavelengths is small. As $T_s$ increases, atmospheric emission reaches saturation at ~320 K in thermal infrared wavelengths, while the absorption in visible and near-infrared wavelengths keeps increasing. This energetic constraint is valid in all the experiments as long as the atmosphere has reached an energy balance (Fig. 2A).

Figures 2B and D suggest that the surface net longwave radiation ($NLW^s$) and surface sensible heat flux ($SH^s$) decrease as a function of $T_s$. The former indicates that the atmosphere emits more thermal radiation downward to the surface, because the atmospheric greenhouse effect increases with $T_s$, primarily due to the water vapor feedback. The latter is primarily caused by the diminishing temperature difference between the surface and near-surface atmosphere as $T_s$ increases. When $T_s$ is higher than ~330 K, $NLW^s$ and $SH^s$ change signs from positive to negative (Fig. 2D), due to the onset of near-surface inversion mainly in the subtropics (*18*). Overall, the change of $SH^s$ is smaller than the other three factors.

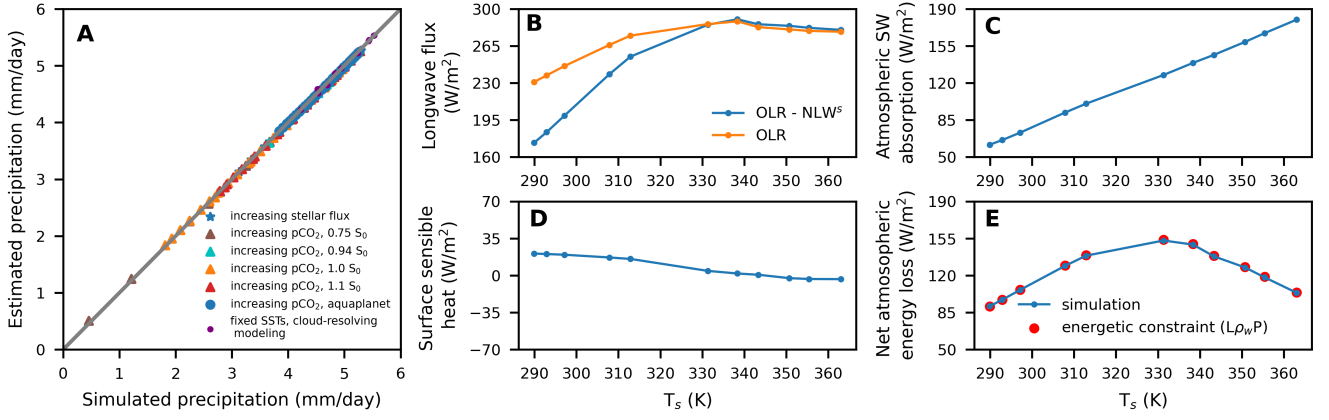

**Fig. 2. Energetic constraint on the mean precipitation.** (**A**) Estimated precipitation based on net atmospheric energy loss (divided by $L\rho_w$) versus simulated precipitation in all the experiments. (**B**) Outgoing longwave radiation at the top of the atmosphere ($OLR$, orange line) and $OLR$ minus net atmospheric thermal emission from the surface ($OLR - NLW^s$, blue line). (**C**) Atmospheric shortwave absorption ($ASW^a$). (**D**) Surface sensible heat the surface ($SH^s$). (**E**) Net atmospheric energy loss (blue line, being equal to $OLR - NLW^s - ASW^a - SH^s$) and the estimated latent heat release related to surface precipitation (red dots). Data in (**B-E**) is from Wolf and Toon (*18*). The strength of mean precipitation is constrained by the ability of net atmospheric radiative cooling.

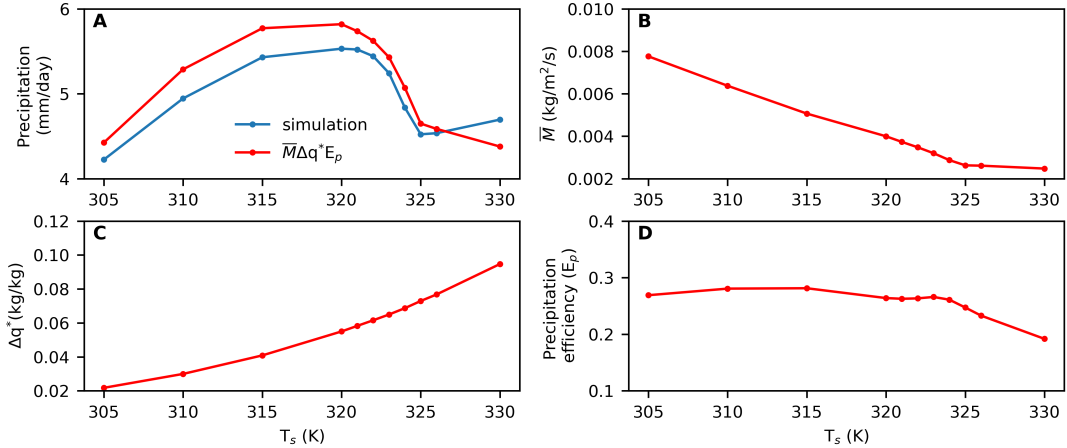

**Fig. 3. Dynamical constraint on the mean precipitation.** (**A**) Simulated precipitation (blue line) and the estimated precipitation based on Eq. (2) (red line). (**B**) Vertically averaged convective mass flux diagnosed from atmospheric energy balance (Eq. (13) in Materials and Methods). (**C**) Saturated water vapor specific humidity difference between cloud base and cloud top. (**D**) Precipitation efficiency. Data is from the cloud-resolving simulations in Seeley and Wordsworth (*22*). The weakening of the mean precipitation under hothouse climates is mainly due to the reduction in convective mass flux.

In the framework of dynamical constraint, water mass conservation suggests that the surface precipitation in an atmospheric column should be approximately equal to the net water vapor flux entering the column from the cloud base (*2*, *15*, *19*, *20*), expressed as follows:

$$P_m \cong \overline{M}\Delta q^* E_p = \overline{M}(q^*_{LCL} - q^*_{ct})E_p, \tag{2}$$

in which $\overline{M}$ is the vertically-averaged convective mass flux between cloud base and cloud top, $q^*_{LCL}$ is the saturated vapor specific humidity at the lifting condensation level (*LCL*), $q^*_{ct}$ is the saturated vapor specific humidity at the cloud top, and $E_p$ is the precipitation efficiency (see Materials and Methods). The value of $q^*_{ct}$, following moist adiabat, is small and negligible in moderate climates but large at hothouse climates. For the cloud-resolving experiments, diagnosed precipitation through Eq. (2) agrees closely with the simulated precipitation with a ~10% difference (Fig. 3A), suggesting that Eq. (2) provides a reasonable estimate to the trend in $P_m$.

As the surface temperature increases, the $P_m$ trend is mainly determined by the increasing rate of $\Delta q^*$ and the decreasing rate of $\overline{M}$, while the change in $E_p$ is relatively small across the temperature range (Fig. 3). As $T_s$ increases, $\Delta q^*$ increases with a rate of about 7%/K following the Clausius-Clapeyron relation (fig. S3B), while $\overline{M}$ decreases. Below the transition temperature (~320 K), $\Delta q^*$ has larger relative changes than $\overline{M}$, so $P_m$ increases with $T_s$. However, when $T_s$ surpasses 320 K, $\overline{M}$ is strongly suppressed, exceeding the effect of the increase in $\Delta q^*$ (fig. S3B), consequently leading to a decrease in $P_m$ with surface warming (fig. S3A). The strong suppression of $\overline{M}$ under warming is mainly caused by the enhancement of atmospheric stratification and the decrease of atmospheric net radiative cooling (*21–23*). Therefore, from a dynamical perspective, surface precipitation weakens in hothouse climates primarily because the convective mass flux is dramatically suppressed. Note that the argument of the dynamical constraint is not fully independent of the energetic constraint; for example, the strength of the net radiative cooling is a key factor for determining the magnitude of $\overline{M}$. Nonetheless, it provides dynamic insights on how the atmospheric motion responds to radiation, subsequently affecting convection and precipitation.

In GCMs, precipitation is comprised of three parts: deep convective precipitation ($P_{deep}$), shallow convective precipitation ($P_{shallow}$), and large-scale precipitation ($P_{large-scale}$). As the surface warms, $P_{deep}$ first increases and then decreases, $P_{shallow}$ continuously increases with $T_s$, and $P_{large-scale}$ monotonically decreases with $T_s$ (fig. S4A). The change of $P_{deep}$ dominates, so the total precipitation changes non-monotonically with $T_s$ with a transition temperature at ~320-330 K. Note that these three types of precipitation are parameterized (*24*) rather than explicitly resolved, therefore their partitions should be model-dependent (*25*). However, the total precipitation should not depend on the parameterization schemes used as it is governed by the energetic constraint shown above.

Above the transition temperature, $P_{deep}$ decreases with $T_s$ because the decreasing rate of deep convective mass flux overweighs the increasing rate of saturation specific humidity (fig. S5, A and B). This mechanism is similar to that found in the cloud-resolving experiments addressed above. The value of $P_{shallow}$ increases monotonically with $T_s$ because the increasing rates of specific humidity and precipitation efficiency outweigh the decreasing rate of shallow convective mass flux (fig. S5, C and D). The tropical Hadley cells and mid-latitudinal synoptic-scale eddies become weaker in a warmer climate (fig. S6), so $P_{large-scale}$ decreases with $T_s$.

Zonal-mean precipitation shows that below the transition temperature, precipitation increases in deep tropics and mid-latitudes (fig. S7). Above the transition temperature, precipitation decreases with $T_s$ in the tropics but increases in polar regions. This trend is consistent with the weakening and expansion of tropical circulation and the poleward shift of the mid-latitude baroclinic zone (fig. S6).

**Applying to other planets**

Besides Earth, our finding is also applicable to other planets, including planets with different stellar spectra (Fig. 1D), slowly rotating planets, and planets with different surface pressures or different gravities (Fig. 4), although the exact values of the maximum precipitation and the transition temperature vary. When changing the stellar spectrum from the Sun to AD Leonis (an M dwarf with an effective temperature of ~3400 K), the maximum precipitation decreases from ~5.5 to 2.5 mm day$^{-1}$ and the transition temperature changes from ~320 to 310 K. This change is caused by a redder spectrum of AD Leonis and the strong absorption of water vapor and $CO_2$ in near-infrared wavelengths (0.7-5.0 μm). Consequently, more stellar energy is absorbed by the atmosphere rather than the surface (*26*, *27*), resulting in weaker precipitation (refer to Eq. (1)) and a lower transition temperature.

For slowly rotating planets, both the maximum precipitation and the transition temperature are lower compared to faster-rotating planets. For example, a planet with a rotation period of 32 Earth days has a maximum precipitation of ~3.5 mm day$^{-1}$ and a transition temperature of ~310 K, while a planet with a rotation period of 256 Earth days has a maximum precipitation of ~2.3 mm day$^{-1}$ and a transition temperature of ~285 K (Fig. 4A). Planetary rotation rate influences the strength of the Coriolis force and the length of day as well as of night. As the rotation period increases, the dayside becomes warmer, leading to a larger diurnal contrast. This enhances night-to-day near-surface convergence, generating more convection during the dayside. Consequently, the dayside water vapor amount and cloud water path increase, resulting in stronger atmospheric shortwave absorption (fig. S8B) and larger planetary albedo (*28–30*). In addition, the small differences between Fig. 4A and Fig. 4B suggest that our conclusion is insensitive to ocean dynamics because Fig. 4A is from coupled atmosphere-ocean experiments within which 3D ocean circulation is properly simulated while Fig. 4B is from atmosphere-only experiments where the ocean is immobile.

As surface pressure increases, the transition temperature remains nearly unchanged, while the maximum precipitation decreases moderately (Fig. 4C). This decrease in maximum precipitation is attributed to two factors. Firstly, with larger surface pressure, pressure broadening and collision-induced continuum absorption increase, resulting in a stronger atmospheric greenhouse effect and a lower $OLR$ under a given $T_s$ (fig. S9A). Additionally, multiple scattering in the atmosphere increases, leading to larger planetary albedo and meanwhile increased shortwave absorption ($ASW^a$) by water vapor (fig. S9B, see also (*31*, *32*)). Both the decreased $OLR$ and the increased $ASW^a$ contribute to a weaker precipitation under a higher surface pressure.

As gravity increases, both the maximum precipitation and transition temperature increase (Fig. 4D). This can be explained by two factors. Firstly, a larger-gravity planet has a lower water vapor column mass (although the vapor pressure remains nearly constant under a given surface temperature), resulting in a weaker greenhouse effect and a larger limiting $OLR$ which is reached at a higher surface temperature (fig. S9D, see also (*14*, *33*, *34*)). Secondly, the dependence of atmospheric shortwave absorption on gravity is tiny (fig. S9E), since a fixed air mass is used in these three experiments.

The precipitation in Fig. 4C and D is derived from 1D radiative transfer modeling and based on the energy conservation analysis (Eq. (1)). The weakening precipitation at high temperatures in this simple framework again confirms the robustness of our conclusion.

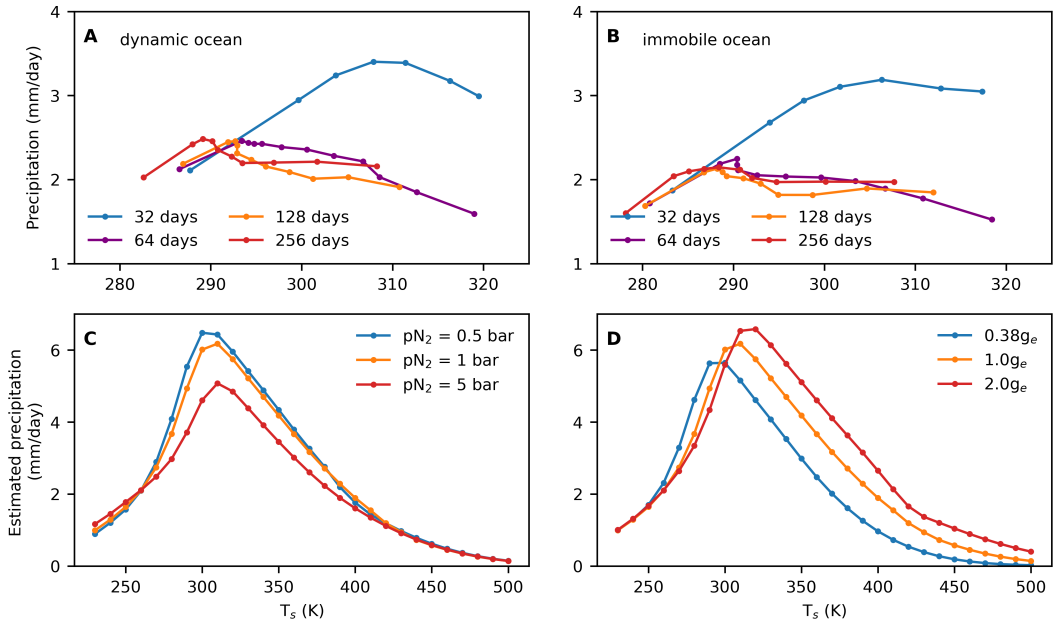

**Fig. 4. Mean precipitation under different planetary parameters.** (**A-B**) 3D global climate simulations from Way et al. (*30*) with planetary rotation periods of 32 (blue), 64 (purple), 128 (orange), and 256 (red) Earth days. The horizontal axis is the global mean surface temperature. The atmosphere is coupled to a dynamic ocean (**A**) or an immobile ocean (**B**). (**C-D**) Estimated mean precipitation based on 1D atmospheric radiative transfer modeling and atmospheric energetic constraint (see Materials and Methods). (**C**) Background air pressure is set to 0.5 bar (blue), 1.0 bar (orange), and 5 bar (red) of $N_2$. (**D**) Surface gravity is set to 0.38 (blue, Mar's value), 1.0 (orange), and 2.0 (red, corresponding to a super-Earth) times Earth's gravity (9.8 m s$^{-2}$) with the same air mass ($\sim 1 \times 10^4$ kg m$^{-2}$). The non-monotonic trend of mean precipitation with $T_s$ does not depend on rotation periods, ocean dynamics, air pressure, or surface gravity.

**Discussion**

The weakening of precipitation under hothouse climates presents a conceptual advance in our understanding of the climate system of Earth as well as of other rocky planets. Contrary to the previous common view that precipitation constantly increases with rising surface temperature, we demonstrate that the increasing rate of precipitation with surface temperature is state-dependent (similar to the climate sensitivity (*35, 36*)). Particularly, the increasing rate approaches zero or even becomes negative under hothouse climates.

High surface temperatures limit planetary habitability, and a weakening hydrologic cycle can lead to further challenges for life. For example, the weakening precipitation may shorten the life span of the future Earth's biosphere under the brightening Sun, compared to previous estimations (*37, 38*). This issue should also be pertinent to Early Earth and Mars with a hot surface or a post-impact steam atmosphere, early

Venus with a high-level insolation, and exoplanets with large amounts of $CO_2$/$CH_4$ or orbiting close to the inner edge of the liquid-water habitable zone (*14*, *18*, *23*, *39*, *40*).

Our discovery challenges the fundamental assumption (precipitation increases with surface temperature) employed in the hypothesis of the carbonate-silicate cycle (*14*, *41*). The weathering rate at high temperatures will be lower than previous anticipation. For instance, in the hothouse climate with high $CO_2$ levels following the melting of a hard snowball Earth that may have occurred during the Neoproterozoic era in ~600-800 million years ago (*42*), the hydrologic cycle strength should be weaker (purple line in Fig. 1B) and the recovery from the post-snowball hot climate should take longer than commonly assumed. Our conclusion confirms and extends previous studies (*6*, *17*). Our results also suggest that in a runaway greenhouse state or an extremely hot climate state (> 500 K), the surface precipitation will approach zero as shown in Fig. 4C and D. Upper-level convection can still occur but all day-side precipitating droplets will re-evaporate at lower levels before reaching the planetary surface (*23*, *43*), similar to the condition observed on modern Venus.

Besides mean precipitation, precipitation variability should also change in hot climates, as suggested by the studies of refs. *22,44-47*. The temporal pattern of precipitation in hothouse climates may transform from quasi-steady to organized episodic deluges, with outbursts of short and heavy rain alternating with several-day dry spells. Future work is required to combine these two aspects (mean and variability) of both regional and global precipitation through a unified theoretical framework.

**Materials and Methods**

We employ simulation data from three atmospheric general circulation models (ExoCAM, ROCKE-3D, and CAM3), two cloud-resolving models (DAM and SAM), and a radiative transfer model (ExoRT) to examine the global-mean or domain-mean precipitation from cold to hot climates. ExoCAM experiments were conducted by Wolf and Toon (*18*), Wolf et al. (*48*), and Zhang et al. (*49*), ROCKE-3D experiments were performed by Way et al. (*30*), DAM experiments were done by Seeley and Wordsworth (*22*), SAM experiments were performed by Dagan et al. (*45*). We utilize their output data in our analysis. Previous investigations primarily focus on surface temperature and clouds (*18*, *30*, *48*, *49*) or the temporal variability of precipitation (*22*, *45*), while our study focuses on the mean precipitation. We conduct experiments using CAM3 and ExoRT (table S1). We employ two distinct methodologies to understand the change in global-mean precipitation: energetic constraint and dynamical constraint.

**ExoCAM**

ExoCAM is a modified version of the Community Earth System Model (CESM) version 1.2. It utilizes the Community Atmosphere Model version 4 (*24*), with a horizontal resolution of 4° × 5° and a finite volume dynamical core. ExoCAM uses a correlated-k two-stream radiative transfer model ExoRT, which is modified to simulate hothouse climates (the hottest air temperature can be up to 500 K) and high-$CO_2$ atmospheres (*50*). The deep convection scheme used in ExoCAM is originally from the parameterization developed by Zhang & McFarlane (*51*) and further modified with the addition of convective momentum transport (*52*) and dilute entraining plumes (*53*, *54*). The scheme is based on a plume ensemble approach, in which an ensemble of convective scale updrafts occurs once the lower atmosphere is conditionally unstable. The shallow convection scheme is treated by the parameterization of Hack (*55*), which can be applied to convection rooted in any layers.

For ExoCAM, we utilize released simulation results of hothouse climates from three published papers: Wolf and Toon (*18*), Wolf et al. (*48*), and Zhang et al. (*49*). Wolf and Toon (*18*) increased the surface temperature by increasing stellar insolation, while Wolf et al. (*48*) and Zhang et al. (*49*) by increasing $CO_2$ concentration. We utilize three sets of experimental data from Wolf et al. (*48*) with different solar constants (0.75, 1.0, and 1.1 times Earth's present-day solar constant, 1360 W m$^{-2}$). The simulations had all reached equilibrium.

Wolf and Toon (*18*) and Wolf et al. (*48*) assumed a 1-bar $N_2$ background, with the addition of $CO_2$ and variable $H_2O$. The total pressure of the atmosphere was the sum of the partial pressures of the three components, $N_2$, $CO_2$, and $H_2O$. Their studies used 45 vertical levels with the model top extending to ~0.2 hPa. Solar spectrum was used. Orbital parameters, such as obliquity, eccentricity, rotation rate, and rotation period, were identical to the present-day Earth. Continental configuration and land coverage were also assumed identical to present-day Earth, except the permanent glacial ice sheets on Antarctica, Greenland, and the Himalayas were replaced by bare soil. A thermodynamic 50-m slab ocean with prescribed internal ocean heat fluxes that mimicked present-day Earth's ocean heat transport was used in their studies. Snow and ice cover were allowed to accumulate once the climates were sufficiently cold.

Zhang et al. (*49*) assumed a 1-bar $N_2$ background with variable $H_2O$. The total pressure of the atmosphere was the sum of the two components, $N_2$ and $H_2O$. The $CO_2$ mixing ratio only affected the radiative transfer scheme and was not included in the total pressure. They used 40 vertical levels with the model top extending to ~1 hPa. Their simulations were conducted with a global ocean surface with no sea ice or continents and with zero ocean heat transport. The stellar insolation (1361 W m$^{-2}$), the stellar spectrum, and the orbital and rotation periods were identical to present-day Earth's values. Both eccentricity and obliquity were set to zero. We used the simulation results of Group InvCM-ClimateSensitivity in their study. This group of simulations assumed fixed global SST with a 1-K increment from 290 to 340 K and the $CO_2$ mixing ratio evolved to equilibrate the system.

**ROCKE-3D**

ROCKE-3D is a generalized version of the Goddard Institute for Space Studies (GISS) ModelE2. It has been developed to allow large ranges of air temperature, air pressure, rotation rate, and atmospheric composition (*56*). ROCKE-3D uses a horizontal resolution of 4° × 5° and 20 vertical levels with the model top extending to 0.1 hPa. For ROCKE-3D, we used the released simulation results from Way et al. (*30*). They used zero obliquity and eccentricity, with land configuration and topography roughly identical to modern Earth. The study assumed a 984-hPa $N_2$ background, with 400 ppmv $CO_2$, 1 ppmv $CH_4$, and variable $H_2O$. The effect of variable $H_2O$ on total air mass was neglected. The atmosphere was coupled to two ocean types: a thermodynamic slab ocean with zero lateral ocean heat transport and a fully coupled dynamic ocean with 10 vertical levels down to 1360 m. In our study, we use the simulation results with long rotation periods of 32, 64, 128, and 256 days, which represent the possible climates of slowly rotating planets like Venus.

## CAM3

CAM3 (Community Atmosphere Model version 3) is the atmospheric component of the Community Climate System Model version 3 (CCSM3) (*57,58*). The model is designed to enable a wide range of spectral resolutions. Here, we use the horizontal resolution of 2.8° × 2.8°. We use 26 vertical levels, with the model top reaching 2 hPa. The atmosphere is coupled to a 50-m slab ocean. To simulate the climate of a post-snowball Earth state, the land configuration is set to be the reconstructed global paleogeography of 635 million years ago (*59*). $O_3$ concentration is set to half of that on modern Earth, $CH_4$ is set to 0.806 ppmv, and $N_2O$ is set to 0.277 ppmv. The solar insolation is set to 1284.98 W $m^{-2}$, 6% lower than the modern Sun. We set the eccentricity to zero, while the obliquity is 23.5°, same as modern Earth. To melt a snowball Earth, $CO_2$ concentration needs to be very high (*60, 61*). A series of $CO_2$ concentrations is tested, 20,000, 40,000, 100,000, 200,000, 300,000, and 400,000 ppmv.

## DAM

DAM is a three-dimensional, finite-volume, fully compressible, non-hydrostatic, cloud-resolving model (*62*). We use the simulation results of hothouse climates from Seeley and Wordsworth (*22*). All the experiments were non-rotating radiative-equilibrium (RCE) simulations. RCE is an idealization of the tropics as a whole, and it is widely used in studying the essential interactions between convection and radiative transfer. Vertical velocity is explicitly resolved in the model. Water vapor condensation occurs when the air parcel reaches saturation. The microphysics scheme is Lin-Lord-Krueger (LLK), which considers six water classes (water, cloud liquid, rain, snow, and graupel). The simulations were conducted on a doubly periodic square domain of 72 km by 72 km, with 140 vertical levels and with free-slip, rigid lids at the top of the model (60 km). The horizontal grid spacings in both directions were 2 km. These grid spacings were high enough to simulate large convective cells such as deep convective plumes and anvil clouds, without using cumulus schemes (*63*). To obtain more realistic radiative heating rates in hothouse climates, Seeley and Wordsworth (*22*) coupled DAM to a line-by-line radiation transfer model (*64*). The diurnal cycle was not included.

Simulation data of both the solar spectrum and an M star spectrum are used in this study. For the solar spectrum experiments, the downward shortwave radiation was 413.13 W $m^{-2}$ at the model top and the solar zenith angle was 43.75°. For the M star spectrum experiments, the downward shortwave radiation was 400 W $m^{-2}$ at the model top and the solar zenith angle was 48.19°. In the G star experiments of DAM, the radiative effects of clouds are considered, but in the M star experiments, no cloud radiative effect is included. However, this does not influence the non-monotonic trend of the mean precipitation as shown in Fig. 1D. The simulations had all reached equilibrium, and the final 100 days of the solar spectrum experiments and the final 50 days of the M star experiments are used here.

## SAM

SAM is a cloud-resolving model using the anelastic dynamic core, documented by Khairoutdinov and Randall (*65*). We use the small-domain RCE simulation results from Dagan et al. (*45*). The simulations were conducted on a doubly periodic square domain of 96 km by 96 km, with 81 vertical levels extending to 33 km. The horizontal resolution was 1 km. Surface temperatures were prescribed, from 295 to 325 K with 2.5 K intervals. Radiation was calculated using CAM3 radiative scheme (*57*). The incoming solar radiation was 551.58 W m$^{-2}$ with a zenith angle of 42.05°. Each simulation was run for 150 days and the last 50 days were used for the analyses.

**Radiative Transfer Modeling**

The one-dimensional (1D) radiative transfer model employed in this study is ExoRT. ExoRT is a two-stream radiative transfer model. It can be used as either an offline model alone or coupled with 3D climate models such as ExoCAM. The latest version n68equiv uses correlated-k coefficients produced with HELIOS-K (*66*). The model uses 68 spectral intervals and 8 gauss points, which can simulate a large range of air temperature (from 100 to 500 K) and air pressure (from 0.00001 to 10 bars).

We do five groups of experiments. In the first three groups, we change the background air pressure (0.5-, 1-, and 5-bar N$_2$) under fixed surface gravity (Earth's value, 9.8 m s$^{-2}$). We use 600 vertical levels, with a grid spacing of 400 m. In the other two groups, we change the surface gravity (0.38- and 2- times Earth's gravity) under fixed dry air mass (~1 × 10$^4$ kg m$^{-2}$). The lower gravity value is the surface gravity of Mars and the higher value is approximately the upper gravity limit of super-Earths. We use 1500 and 300 vertical levels for the 0.38- and 2-times Earth's gravity experiments, respectively.

We specify the surface temperature for each group, from 230 to 500 K with 10-K increments. The surface albedo is set to 0.22, considering the effect of clouds and sea ice. The solar constant is set to 680 W m$^{-2}$ with a zenith angle of 60°. The atmosphere is comprised of N$_2$ and H$_2$O. We assume the air temperature follows H$_2$O moist adiabatic profile in the troposphere and the stratosphere is set to isothermal at 200 K. We consider water vapor to be saturated, with its concentration fixed according to the air temperature profile. The inverse moist adiabatic lapse rate is calculated by

$$\frac{d \ln P}{d \ln T} = \frac{P_n}{P} \frac{d \ln P_n}{d \ln T} + \frac{P_c}{P} \frac{d \ln P_c}{d \ln T}, \qquad (3)$$

where $P_c$ and $P_n$ are the partial pressures of H$_2$O and N$_2$, $P$ is the total pressure, and $T$ is the air temperature. We use the hydrostatic equation to convert the pressure coordinate to the height coordinate. The first term on the right-hand side (RHS) of Eq. (3) represents the inverse dry adiabatic lapse rate, which, following Pierrehumbert (*14*), can be expressed as:

$$\frac{d \ln P_n}{d \ln T} = \frac{c_{pn}}{R_n} \times \frac{1 + \left(\frac{c_{pc}}{c_{pn}} + \left(\frac{L}{R_c T} - 1\right) \frac{L}{c_{pn} T}\right) \alpha_c}{1 + \frac{L}{R_n T} \alpha_c}. \qquad (4)$$

Here, $c_{pn}$ and $c_{pc}$ are the specific heat capacities of N$_2$ and H$_2$O, $R_n$ and $R_c$ are the specific gas constants of N$_2$ and H$_2$O, and $\alpha_c$ is the mass mixing ratio of H$_2$O. We assume that both N$_2$ and H$_2$O behave as ideal gases. The second term on the RHS of Eq. (3) follows the Clausius-Clapeyron relation

$$\frac{d \ln P_c}{d \ln T} = \frac{L}{R_c T}, \qquad (5)$$

where $L$ is the latent heat vaporization of water vapor. To obtain more precise results, we take into account the dependence of specific heat capacity ($c_p$) and latent heat of vaporization ($L$) on air temperature. The specific heat capacity function with respect to temperature can be described by the Shomate equation (*14*)

$$c_p = A + B \times \frac{T}{1000} + C \times \left(\frac{T}{1000}\right)^2 + D \times \left(\frac{T}{1000}\right)^3 + E \times \left(\frac{T}{1000}\right)^{-2} \ J\ kg^{-1}\ K^{-1}. \tag{6}$$

For N$_2$, this equation is valid from 300 to 500 K, with coefficients A, B, C, D, and E being as 931.857, 293.529, -70.576, 5.688, and 1.587, respectively (*14*). Below 300 K, the specific heat capacity of N$_2$ is set to 1037 J kg$^{-1}$ K$^{-1}$. For H$_2$O, this equation is valid from 273 to 1800 K, with coefficients of 161.778, 379.584, 377.413, -140.804, and 4.563, respectively (*67*). Below 273 K, the specific heat capacity is set to 1847 J kg$^{-1}$ K$^{-1}$. The latent heat of vaporization changes with temperature following Kasting (*68*),

$$\frac{dL}{dT} = c_{pv} - c_w \approx 2230 \ J\ kg^{-1}\ K^{-1}, \tag{7}$$

where $c_w$ is the specific heat capacity of liquid water. Thus, $L$ can expressed as $2.5 \times 10^6 - 2230 \times (T - 273.15)\ J\ kg^{-1}$. The saturated water vapor pressure ($P_c$) is given by experimental data as (*69*)

$$P_c = \frac{e^{77.3450 + 0.0057T - 7235/T}}{T^{8.2}}. \tag{8}$$

The estimated precipitation derived from the 1D radiative transfer modeling (Fig. 4, C and D) is determined by the atmospheric energy constraint, with surface sensible heat flux assumed to be zero (following Eq. (1) with $SH^s = 0$). Clouds are not included in the radiative transfer modeling.

**Energetic Constraint on Precipitation**

The energetic constraint is expressed as Eq. (1) and illustrated in fig. S1A. Horizontal and vertical heat and water transports are not explicitly included in the equation. In experiments where the atmosphere is interactively coupled to the surface, both the atmospheric energy budget and the surface energy budget are closed once the experiment has reached equilibrium. In experiments where the surface temperature is fixed, the atmospheric energy budget is closed but not the surface energy budget. In our calculations, we use the atmospheric energy budget to estimate the strength of mean precipitation.

In our analysis, both cloud shortwave and longwave radiative effects are considered unless stated otherwise. Cloud shortwave radiative effect tends to increase planetary albedo thereby reducing shortwave absorption in the atmosphere and at the surface. In addition, cloud particles can absorb near-infrared radiation, leading to a stronger atmospheric shortwave absorption. However, this effect typically has a smaller magnitude compared to water vapor absorption. Consequently, the cloud shortwave radiative effect is not typically the dominant term in the atmospheric energy budget. On the other hand, cloud longwave radiative effect tends to reduce outgoing longwave radiation to space, resulting in a weakening precipitation.

Different GCMs employ different parameterization schemes for clouds, so the simulated values in the cloud longwave radiative effect should be model-dependent. The cloud-resolving model DAM and SAM can resolve clouds at gird scales of 4 km² and 1 km² respectively, but cloud microphysics (such as cloud particle sizes and cloud droplet falling speeds) still require parameterization. For experiments employed in this study, the changes of cloud longwave radiative effect are limited to 10-20 W m$^{-2}$. This limitation is likely due to the nearly constant cloud top temperature (~200-220 K), proposed by the fixed anvil temperature (FAT) hypothesis (*70*).

Given that an energy of 10 W m$^{-2}$ corresponds to a mean precipitation change of approximately 0.35 mm day$^{-1}$, such variations are not potent enough to substantially alter the overall trend of precipitation as a function of surface temperature. This suggests that our results are robust to cloud and convection parameterization as shown in Figs. 1 and 4. There is only one exception in our experiments: the blue line in Fig. 1D shows that the mean precipitation at 330 K appears somewhat stronger than that at 325 K. This is primarily attributed to the reduction of the cloud longwave radiative effect from 35.2 W m$^{-2}$ at 325 K to 31.9 W m$^{-2}$ at 330 K.

**Dynamical Constraint on Precipitation in DAM**

In the RCE simulations using the cloud-resolving model DAM, the estimation of precipitation can be expressed as Eq. (2). This equation, popularized by Held and Soden (*2*), has been widely employed in numerous studies for diagnosing or analyzing precipitation (e.g., *16*, *20*, *71*). We derive Eq. (2) by transforming the original form of Eq. (12) presented in Jeevanjee (*20*):

$$P_m = E_p \int_{z_{LCL}}^{z_{ct}} (-M \frac{dq_v^*}{dz} - \epsilon M(1-RH)q_v^*)dz. \tag{9}$$

Here, $E_p$ is precipitation efficiency, $M$ is the convective mass flux, $q_v^*$ is the saturated water vapor specific humidity, $z_{LCL}$ and $z_{ct}$ is the height of the lifting condensation level and cloud top, $\epsilon$ is the fractional entrainment per unit distance (m$^{-1}$), and $RH$ is relative humidity. Romp (*72*) suggests a characterized $RH$ value of ~75-80% in the tropics. For simplicity, we assume $RH$ is close to one, allowing us to approximate:

$$P_m \cong E_p \int_{z_{cb}}^{z_{ct}} -M \frac{dq_v^*}{dz} dz. \tag{10}$$

By assuming $RH$ close to one, we neglect the suppression on condensation resulting from the entrainment of drier environmental air into the convective mass. Consequently, this results in an overestimation of mean precipitation. Since M and $q_v^*$ are nearly independent and the change of M with height is relatively small compared to the change of $q_v^*$, we can express the above equation as:

$$P_m \cong -E_p \frac{\sum(M\delta z)}{z_{cb} - z_{ct}} \int_{z_{cb}}^{z_{ct}} \frac{dq_v^*}{dz} dz = E_p \frac{\sum(M\delta z)}{z_{cb} - z_{ct}} (q_{LCL}^* - q_{ct}^*). \tag{11}$$

In this equation, $\frac{\sum M\delta z}{z_{cb}-z_{ct}}$ represents the vertically average convective mass flux, which accounts for the effects of entrainment and detrainment. By using a more concise term $\bar{M}$ to replace $\frac{\sum M\delta z}{z_{cb}-z_{ct}}$, the equation can be rewritten as Eq. (2): $P \cong \bar{M}\Delta q^* E_p = \bar{M}(q^*_{LCL} - q^*_{ct})E_p$.

In Eq. (2), $\bar{M}\Delta q^*$ represents the total water vapor flux in the domain that has the potential to generate precipitation. In moderate climates, $q^*_{ct}$ is typically close to zero and can be neglected. However, in very hot climates, $q^*_{ct}$ can reach values around 10$^{-1}$, making it non-negligible (fig. S10F-J). Moreover, not all condensation can transform to rain and reach the surface, as some may reevaporate when falls into a sub-saturated or cloud-free layer. Thus, precipitation efficiency ($E_p$), the rate of net condensation to gross condensation in the whole air column, is included in the estimation. Equation (2) allows us to account for the impacts of entrainment and detrainment without explicitly constraining them.

In the cloud-resolving modeling, the convective updraft grid cells can be determined as the cloud condensation mixing ratio exceeding a threshold (e.g., 10$^{-5}$) and the vertical velocity being greater than a critical value (e.g., 1 m s$^{-1}$). Subsequently, the convective mass flux can be obtained from $M = \rho_a w_{up} \sigma_{up}$, in which $\rho_a$ is the air density, $w_{up}$ is the vertical velocity averaged over the whole updraft grid cells, and $\sigma_{up}$ is the fraction area of the updraft grid cells over the entire domain.

Unfortunately, the released data by Seeley and Wordsworth (*22*) did not include the convective mass flux. Consequently, we diagnose $M$ based on mass conservation and atmospheric energy balance. According to the principle of mass conservation, convective mass flux should have an equal value and be opposite to the subsidence mass flux, $\rho_a w_{sub}$, where $w_{sub}$ ($< 0$) is the averaged vertical velocity of the subsidence region. In the clear-sky subsidence region, it is the radiative cooling ($Q_{rad}$) and evaporating cooling ($Q_{evap}$) balanced by the dynamic heating ($Q_{dyn}$) of descending parcels by expansion, written as $Q_{dyn} = -(Q_{rad} + Q_{evap})$ (*73*, *74*). Applying the weak temperature gradient approximation (*75*), $Q_{dyn}$ in the subsidence region is approximately equal to $-\frac{w_{sub}}{c_p}\frac{\partial}{\partial z}(c_p T + gz)$, where $c_p$ is the specific heat capacity of air at constant pressure, $g$ is surface gravity, and $z$ is height (*76*). Thus, the subsidence velocity can be approximated as

$$w_{sub} \cong \frac{-(Q_{rad} + Q_{evap})}{\frac{1}{c_p}\frac{\partial}{\partial z}(c_p T + gz)} = \frac{Q_{rad} + Q_{evap}}{\Gamma_d - \Gamma}, \quad (12)$$

where $\Gamma_d$ and $\Gamma$ are the dry adiabatic lapse rate and environmental lapse rate, respectively. Consequently, the convective mass flux can be expressed as

$$M \cong -\rho_a w_{sub} \cong -\rho_a \frac{(Q_{rad} + Q_{evap})}{\Gamma_d - \Gamma}. \quad (13)$$

Since the convective updraft region typically occupies a small fraction (~1%) of the entire domain, the subsidence area fraction is close to 1 (*20*). Therefore, we can calculate the convective mass flux profile using domain-mean radiative cooling, evaporating cooling, and lapse rate. This approach was verified in Jeevanjee (*20*) by comparing the convective mass flux profile with the $-\rho_a w_{sub}$ profile (see Figure 3 in ref. (*20*)). In our study, we employ the same method to diagnose the convective mass flux ($M$), and the vertically averaged $\bar{M}$ is calculated as $-\overline{\rho_a w_{sub}}$.

To quantitatively assess the effects of the three factors ($\bar{M}$, $\Delta q^*$, and $E_p$) on the trend of mean precipitation, we use a fractional type of Eq. (2), given by:

$$\frac{\delta P}{P \delta T_s} \cong \frac{\delta \bar{M}}{\bar{M} \delta T_s} + \frac{\delta \Delta q^*}{\Delta q^* \delta T_s} + \frac{\delta E_p}{E_p \delta T_s} + res. \tag{14}$$

The first three terms are the fractional changes of $\bar{M}$, $\Delta q^*$, and $E_p$, respectively, and the fourth term is the residual term ($res$), which comprises four small cross-section terms:

$$res \equiv \frac{\delta \bar{M} \delta \Delta q^*}{\bar{M} \Delta q^* \delta T_s} + \frac{\delta E_p \delta \Delta q^*}{E_p \Delta q^* \delta T_s} + \frac{\delta E_p \delta \bar{M}}{E_p \bar{M} \delta T_s} + \frac{\delta E_p \delta \bar{M} \delta \Delta q^*}{P \delta T_s}. \tag{15}$$

**Dynamical Constraint on Precipitation in ExoCAM**

The precipitation in the general circulation model ExoCAM consists of three components: deep convective precipitation ($P_{deep}$), shallow convective precipitation ($P_{shallow}$), and large-scale (non-convective and stratiform) precipitation ($P_{large-scale}$). We employ a similar equation to diagnose $P_{deep}$ as that used in the cloud-resolving model by

$$P_{deep} \cong \overline{M_d} \Delta q^* E_{pd} = \overline{M_d}(q^*_{LCL} - q^*_{ct}) E_{pd}, \tag{16}$$

where $\overline{M_d}$ is the vertically averaged deep convective mass flux (including both dry air and water vapor) between the cloud base and cloud top, $q^*_{LCL}$ is the saturated water vapor specific humidity at the lifting condensation level ($LCL$), $q^*_{ct}$ is the saturated vapor specific humidity at the cloud top, and $E_{pd}$ is the deep convective precipitation efficiency calculated as the ratio of net deep convective condensation to gross condensation in the deep convective scheme. Similar to Eq. (14), the fractional change of $P_{deep}$ can be derived and the results are shown in fig. S5.

Similarly, the shallow convective precipitation ($P_{shallow}$) can be diagnosed by

$$P_{shallow} \cong W_{shallow} E_{ps} = \overline{M_s} \Delta q_{eq} E_{ps}. \tag{17}$$

Here, $W_{shallow}$ is the total shallow convective water flux, $\overline{M_s}$ is the averaged shallow convective mass flux, $\Delta q_{eq}$ is the equivalent water vapor specific humidity difference between cloud base and cloud top for shallow convection, and $E_{ps}$ is the shallow convective precipitation efficiency. Similar to Eq. (14), the fractional change of $P_{shallow}$ can be expressed by:

$$\frac{\delta P_{shallow}}{P_{shallow}\delta T_s} \cong \frac{\delta \overline{M_s}}{\overline{M_s}\delta T_s} + \frac{\delta \Delta q_{eq}}{\Delta q_{eq}\delta T_s} + \frac{\delta E_{ps}}{E_{ps}\delta T_s} + res. \tag{18}$$

The residual term ($res$) includes four small cross-section parts: $\frac{\delta \overline{M_s}\delta \Delta q_{eq}}{\overline{M_s}\Delta q_{eq}\delta T_s}$, $\frac{\delta E_{ps}\delta \Delta q_{eq}}{E_{ps}\Delta q_{eq}\delta T_s}$, $\frac{\delta E_{ps}\delta \overline{M_s}}{E_{ps}M_s\delta T_s}$, and $\frac{\delta E_{ps}\delta \overline{M_s}\delta \Delta q_{eq}}{P_{shallow}\delta T_s}$. The values of $W_{shallow}$ and $\overline{M_s}$ are available in the model output, but $\Delta q_{eq}$ is not. Therefore, we calculate $\Delta q_{eq}$ as $W_{shallow}/\overline{M_s}$, representing an equivalent value of water vapor specific humidity in shallow convection. $E_{ps}$ is calculated as the ratio of net shallow convective condensation to the sum of gross condensation and the amount of liquid water detrained to neighboring grid cells in the shallow convective scheme. The inclusion of detrained liquid water is essential as liquid water may disperse into the environment within the shallow convective scheme (*24*). Note that shallow convection can initiate at any level within the model once the necessary condition is met (*24*), so the diagnosis is more complex than deep convection.

Large-scale precipitation ($P_{large-scale}$) is parameterized within each grid cell of the model. It is very difficult (if not impossible) to diagnose because many processes should be involved, such as the large-scale (resolvable) ascending in grid cells, the horizontal divergence of water vapor, and the detrained liquid water from convective columns. Rather than directly assessing $P_{large-scale}$, we address its trend through analyzing the changes in the atmospheric overturning streamfunction and mid-latitude eddies (see fig. S6).

**Acknowledgments**: We thank E. T. Wolf for releasing ExoRT. We appreciate Eric T. Wolf, Michael J. Way, Yixiao Zhang, Jacob T. Seeley, and Guy Dagan for generously sharing their data.

**Funding:** This work is supported by the Natural Science Foundation of China (NSFC) under grant nos. 42275134, 41888101, 42075046, and 42161144011

**Author contributions:** J.Y. led this project. J.Y. and J.L. designed and did the experiments. J.L. did the analyses and plotted the figures. Y.H., J.Y., G.C., F.D., and J.L. discussed the results. J.Y. and J.L. wrote the draft, and all authors improved the manuscript.

**Competing interests:** The authors declare no competing interests.

**Data and materials availability:** All data needed to evaluate the conclusions in the paper are present in the paper and/or the Supplementary Materials. The model CAM3 and ExoRT are publicly available at https://www.cesm.ucar.edu/models/cam and https://github.com/storyofthewolf/ExoRT. Data from Wolf and Toon (*18*), Wolf et al. (*48*), Zhang et al. (*49*), Way et al. (*30*), Seeley and Wordsworth (*22*), Dagan et al. (*45*) can be obtained from https://archive.org/details/TheEvolutionOfHabitableClimatesUnderTheBrighteningSun, https://archive.org/details/EvaluatingClimateSensitivityToCO2AcrossEarthsHistory_201809, https://archive.org/details/Climates_of_Warm_Earth_like_Planets, https://knowledge.uchicago.edu/record/3417, https://zenodo.org/records/5636455, and https://zenodo.org/records/8054055. Simulation data from CAM3 and ExoRT, as well as the codes used for analysis and figure plotting in this study, are stored and accessible at: https://zenodo.org/records/10786266.

**Supplementary Materials**

Figs. S1 to S10

Table S1

# Science Advances

## AAAS

Supplementary Materials for

**Hydrologic Cycle Weakening in Hothouse Climates**

Jiachen Liu *et al.*

Corresponding author: Jun Yang, email: junyang@pku.edu.cn

**The PDF file includes:**

Figs. S1 to S10

Table S1

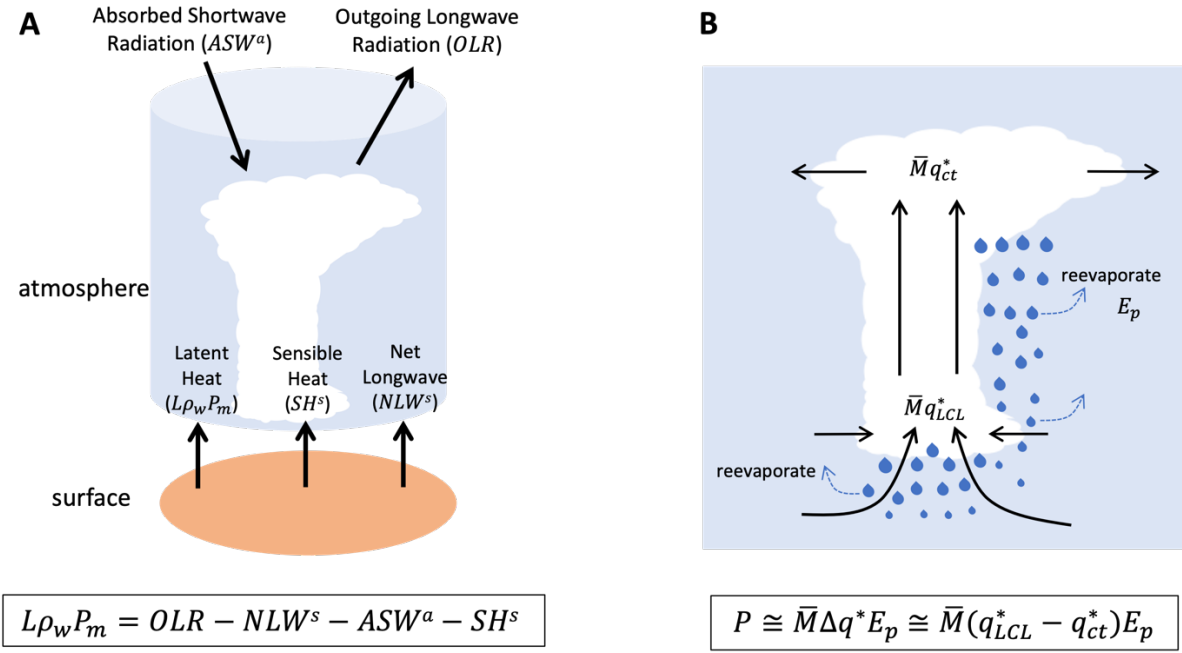

**Fig. S1. Schematic of energetic constraint (A) and dynamical constraint (B) of precipitation.** In energetic constraint, the mean precipitation strength is determined by net atmospheric energy loss. In dynamical constraint, the mean precipitation strength is mainly determined by convective mass flux, specific humidity difference between the cloud base and cloud top, and precipitation efficiency.

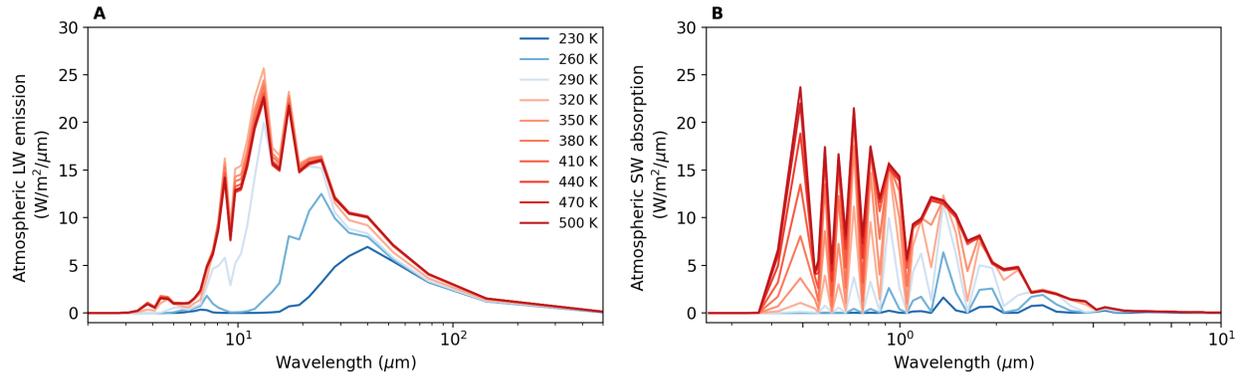

**Fig. S2. Spectral distribution of (A) atmospheric longwave emission ($OLR - NLW^s$) and (B) atmospheric shortwave absorption ($ASW^a$).** Results are from 1D radiative transfer calculations using the model ExoRT with 1 bar background $N_2$ and Earth's gravity. The model uses 68 spectral intervals. Different lines are for different surface temperatures ranging from 230 to 500 K with intervals of 30 K.

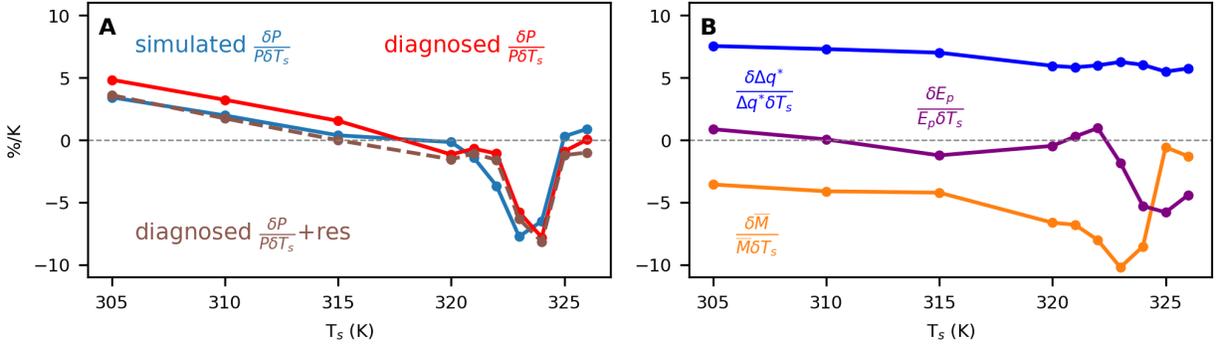

**Fig. S3. Fractional change of precipitation and each contributor over $T_S$ in DAM.** (**A**) Simulated fractional change of precipitation over $T_S$ (blue), diagnosed fractional change of precipitation over $T_S$ by $\bar{M}$, $\Delta q^*$, and $E_p$ (red) based on Eq. (14), and diagnosed fractional change of precipitation involving the residual term (brown) based on Eq. (15). (**B**) Fractional change of $\bar{M}$ (orange), $\Delta q^*$ (dark blue), and $E_p$ (purple). Data is from the cloud-resolving simulations in Seeley and Wordsworth (*22*).

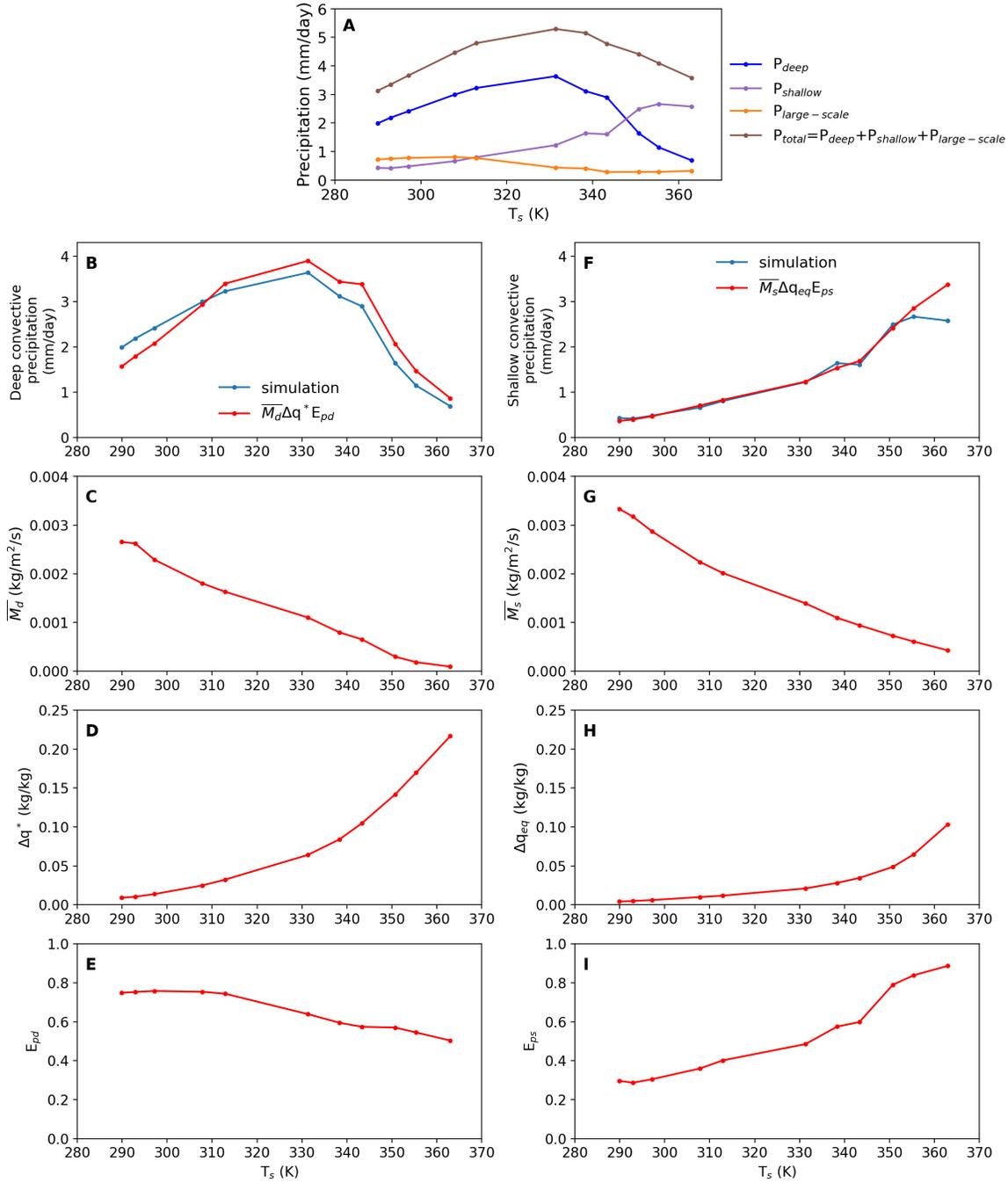

**Fig. S4. Dynamical constraint on the precipitation in ExoCAM.** (**A**) Simulated total (brown), deep convective (blue), shallow convective (purple), and large-scale (orange) precipitation. (**B-E**) are for deep convection, and (**F-I**) are for shallow convection. (**B**) Simulated (blue) and estimated (red) deep convective precipitation based on deep convective mass flux (Eq. (16)). (**C**) Vertically-averaged deep convective mass flux. (**D**) Saturated water vapor specific humidity difference between cloud base and cloud top. (**E**) Deep convective precipitation efficiency. (**F-I**) are the same as (**B-E**) but for shallow convection. The simulation data is from Wolf and Toon (*18*).

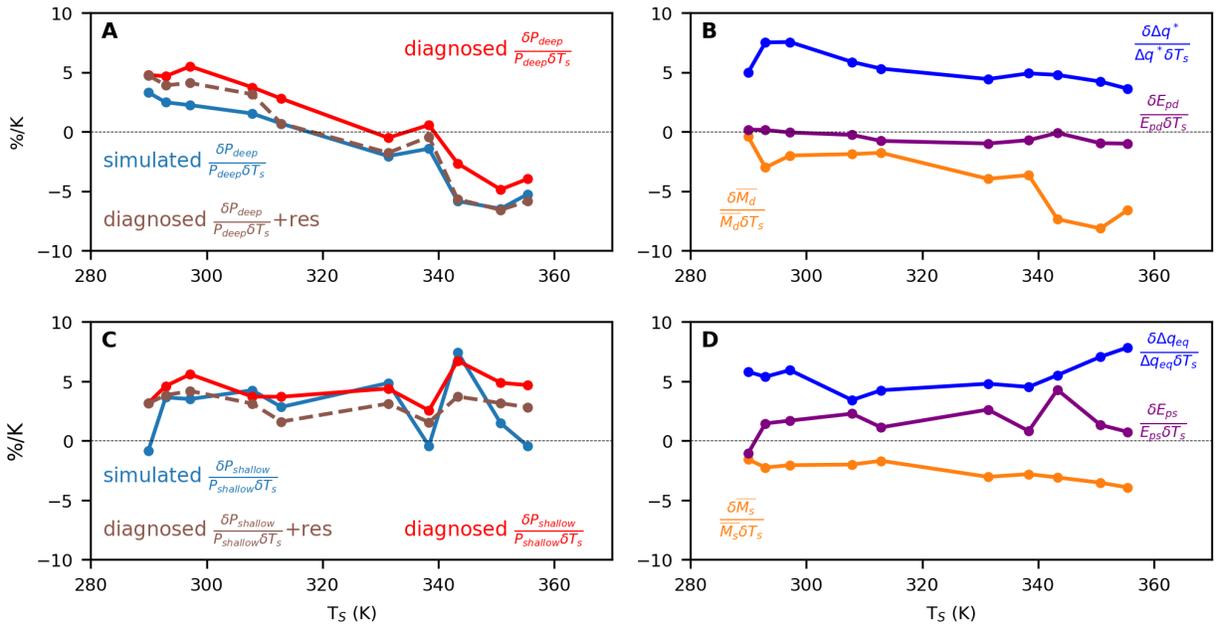

**Fig. S5. Fractional change of convective precipitations and the contributors.** (**A-B**) are the same as fig. S3, but for deep convection in the model ExoCAM. (**C-D**) are the same as (**A-B**) but for shallow convection in the model. The simulation data is from Wolf and Toon (*18*).

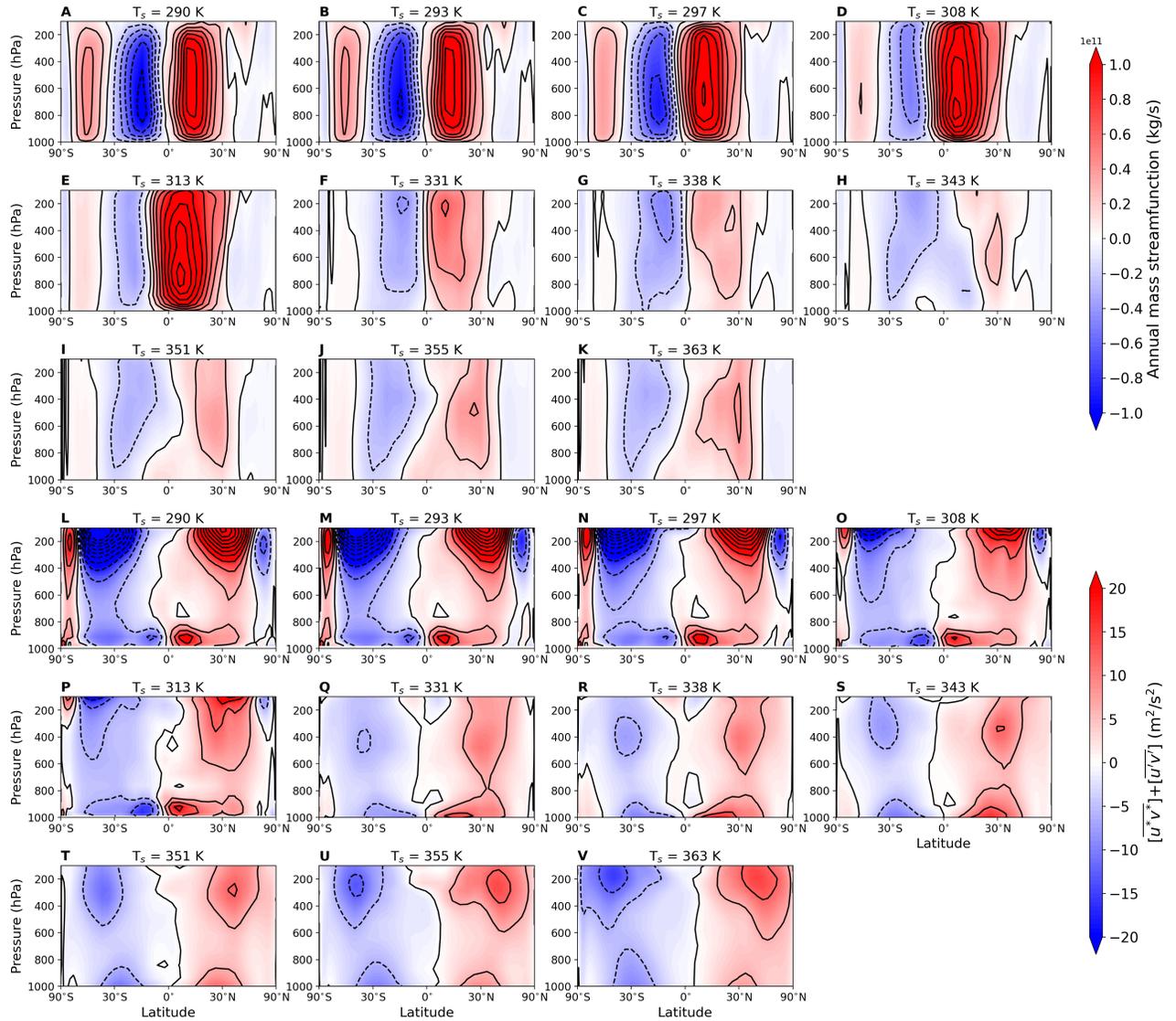

**Fig. S6. Annual-mean meridional mass stream function (A-K) and eddy momentum flux (L-V) under different $T_S$ in ExoCAM.** For the mass stream function, the lower and upper bounds are $-10^{11}$ and $10^{11}$ kg s$^{-1}$ with intervals of $2 \times 10^{10}$ kg s$^{-1}$ for both the color bar and contour lines. For the eddy momentum flux, the lower and upper bounds are -20 and 20 m$^2$ s$^{-2}$ for the color bar and -60 and 60 m$^2$ s$^{-2}$ for contour lines with intervals of 6 m$^2$ s$^{-2}$. The simulation data is from Wolf and Toon (*18*).

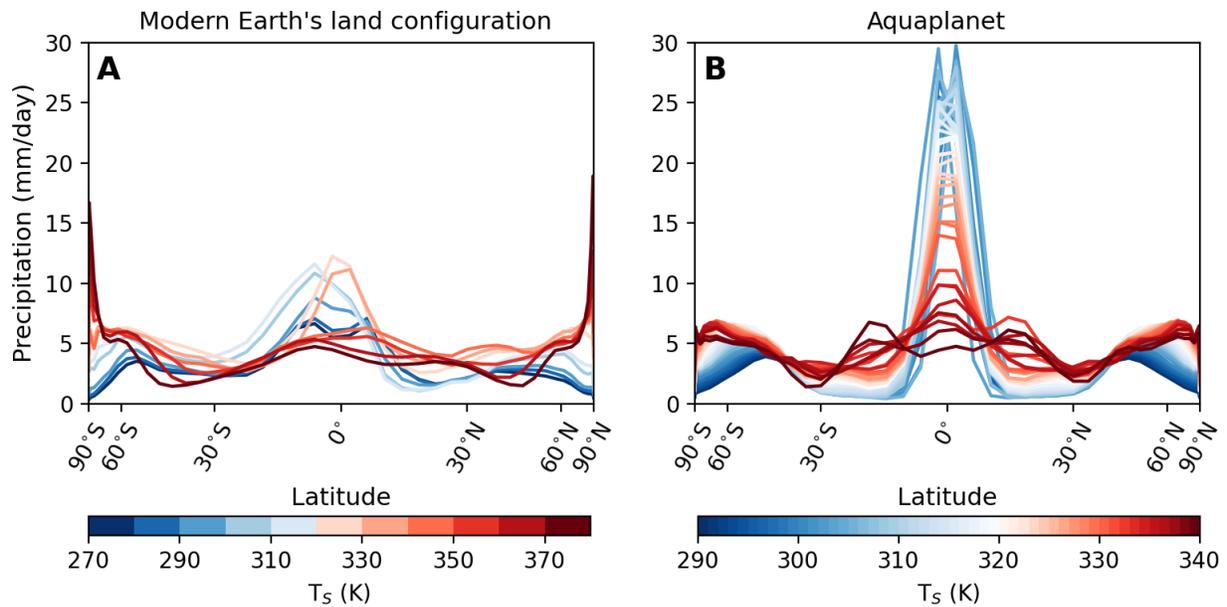

**Fig. S7. Zonal-mean surface precipitation under different $T_S$ in ExoCAM. (A)** The simulation data is from Wolf and Toon (*18*) with modern Earth's land configuration. **(B)** The simulation data is from Zhang et al. (*49*) with a global ocean surface (an aqua-planet).

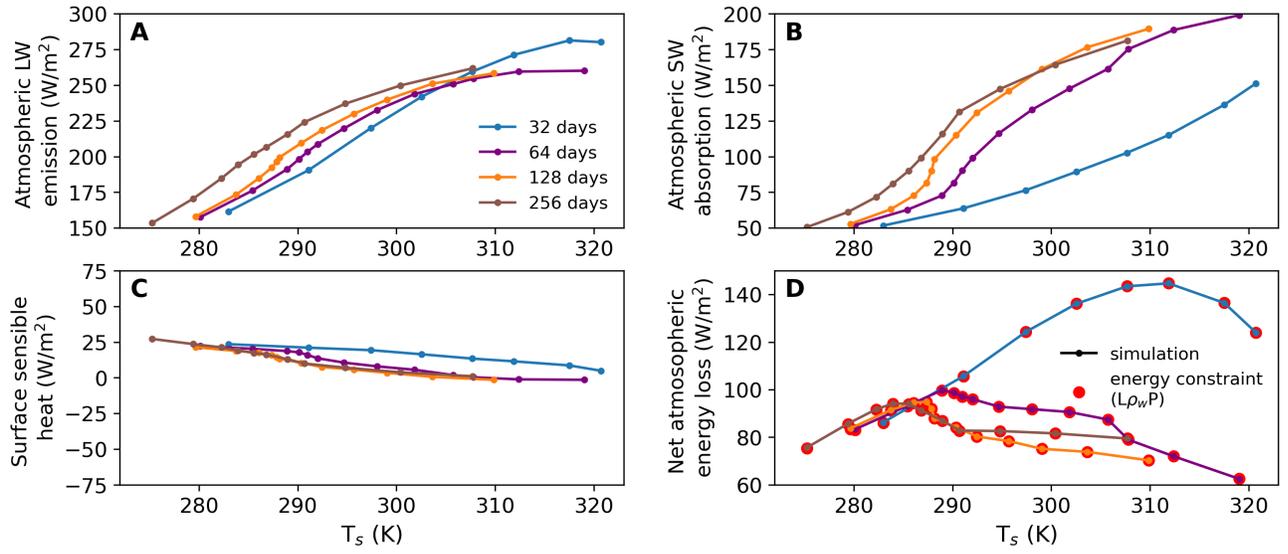

**Fig. S8. Energetic constraint on surface precipitation in the simulations of slow-rotating planets.** The rotation periods are 32 (blue), 64 (purple), 128 (orange), and 256 (brown) days. (**A**) Net atmospheric longwave emission ($OLR - NLW^s$). (**B**) Atmospheric shortwave absorption ($ASW^a$). (**C**) Surface sensible heat ($SH^s$). (**D**) Net atmospheric energy loss (color lines, being equal to $OLR - NLW^s - ASW^a - SH^s$) and the estimated latent heat release based on surface precipitation (red dots). Data is from global simulations coupled to a dynamic ocean using ROCKE-3D in Way et al. (*30*).

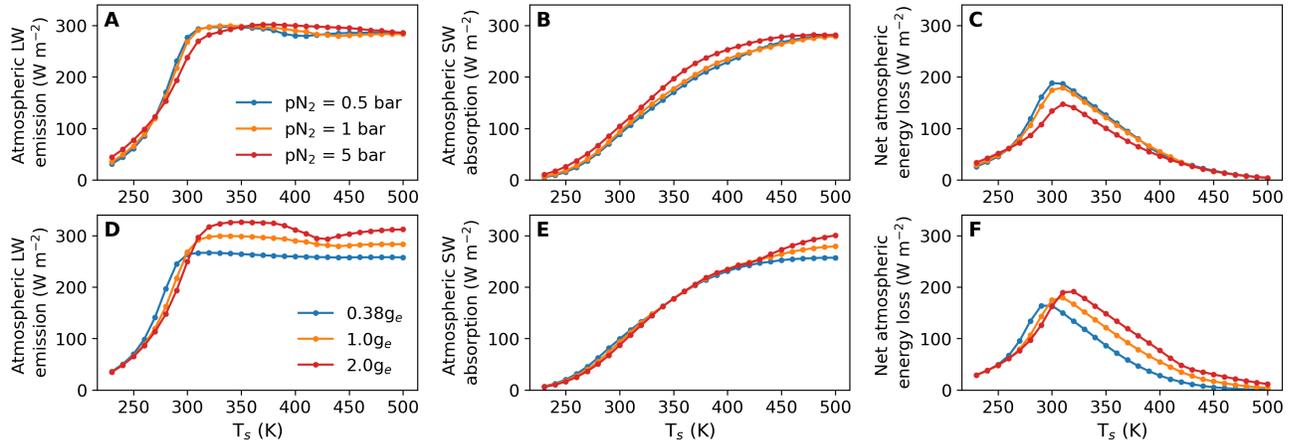

**Fig. S9. Energetic constraint on mean precipitation in radiative transfer modeling.** From (**A**) to (**C**), simulations are under different background air pressures: 0.5 (blue), 1 (orange), and 5 bar (red) $N_2$. From (**D**) to (**F**), under different surface gravities: 0.38 (blue), 1.0 (orange), and 2.0 (red) times Earth's gravity ($g_e$) with the same air mass ($\sim 1 \times 10^4$ kg m$^{-2}$). (**A** and **D**) Atmospheric longwave emission ($OLR - NLW^s$). (**B** and **E**) Atmospheric shortwave absorption ($ASW^a$). (**C** and **F**) Net atmospheric energy loss ($OLR - NLW^s - ASW^a - SH^s$). Results are from 1D radiative transfer calculations using the model ExoRT.

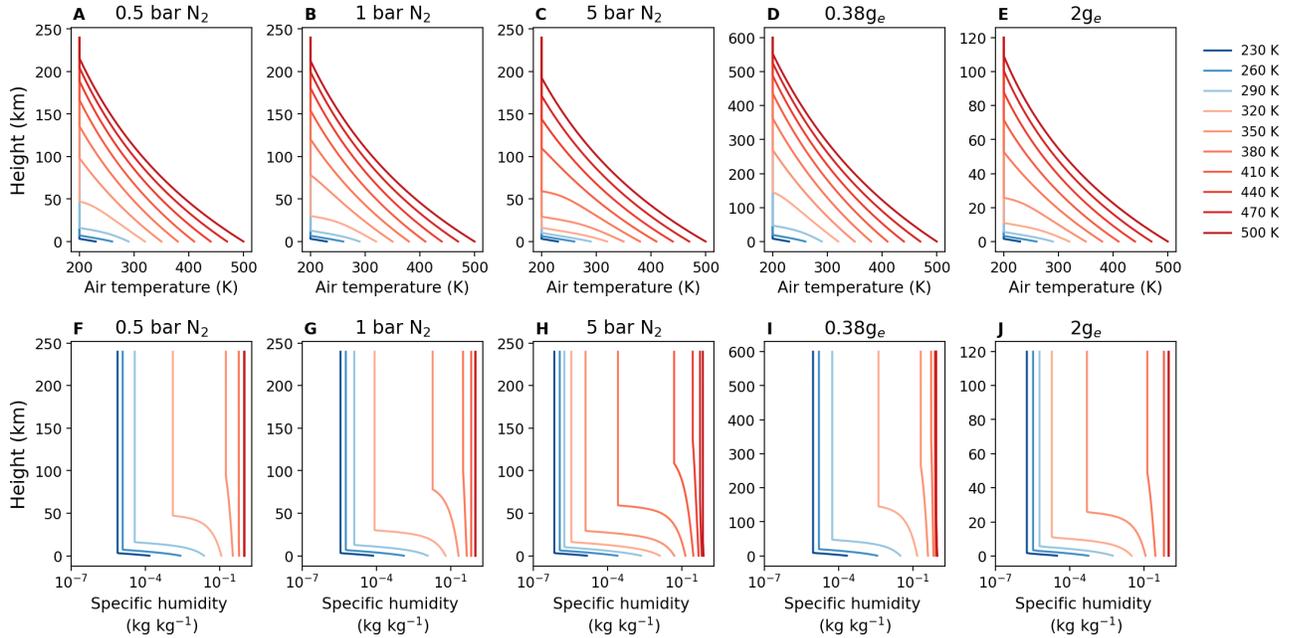

**Fig. S10. Vertical profiles of air temperature and specific humidity in the 1D radiative transfer modeling.** From (**A**) to (**E**), air temperature profiles for simulations of 0.5 bar $N_2$ with Earth's gravity ($1g_e$), 1 bar $N_2$ with $1g_e$, 5 bar $N_2$ with $1g_e$, 1 bar $N_2$ with $0.38g_e$, and 1 bar $N_2$ with $2g_e$. From (**F**) to (**J**), specific humidity profiles for simulations of 0.5 bar $N_2$ with $1g_e$, 1 bar $N_2$ with $1g_e$, 5 bar $N_2$ with $1g_e$, 0.38 bar $N_2$ with $0.38g_e$, and 2 bar $N_2$ with $2g_e$. The troposphere is assumed to be saturated everywhere in all the experiments. In each panel, the lines are for different surface temperatures ranging from 230 to 500 K with intervals of 30 K. For the analytic solution of the temperature profiles, please see Eq. (3) in Materials and Methods.

**Table S1. Summary of the simulation data used in the study**

| Model | Experimental Designs | $T_s$ range | Data source |
|---|---|---|---|
| ExoCAM | • Modern Earth's continental distribution<br>• Increasing stellar flux | 290 to 365 K | Wolf and Toon (*18*) |
| ExoCAM | • Modern Earth's continental distribution<br>• Three different stellar fluxes: 0.75, 1.0, & 1.1 present solar constant<br>• Increasing $CO_2$ concentration | 270 to 380 K | Wolf et al. (*48*) |
| CAM3 | • Post-snowball Earth ~630 million years ago<br>• Reconstructed global paleogeography<br>• Increasing $CO_2$ concentration under 0.94 present solar constant | 300 to 350 K | This study |
| ExoCAM | • Aqua-planet with no continent<br>• Increasing $CO_2$ concentration under present solar constant<br>• Fixed global SST | 290 to 340 K | Zhang et al. (*49*) |
| DAM | • Cloud-resolving small-domain simulations<br>• Model resolution: 2 km × 2 km<br>• Domain size: 72 km × 72 km<br>• Fixed SST with solar or M star spectrum | 305 to 330 K | Seeley and Wordsworth (*22*) |
| SAM | • Cloud-resolving small-domain simulations<br>• Model resolution: 1 km × 1 km<br>• Domain size: 96 km × 96 km<br>• Fixed SST with solar spectrum | 295 to 325 K | Dagan et al. (*45*) |
| ROCKE-3D | • Modern Earth's continental distribution<br>• Rotation periods: 32, 64, 128, and 256 Earth days<br>• Increasing stellar flux | 260 to 320 K | Way et al. (*30*) |
| ExoRT | • 1D radiative transfer model<br>• Fixed surface temperature: from 230 to 500 K<br>• Fixed air temperature and water vapor profiles<br>• Clouds are not included<br>• Different values of gravity: 0.38, 1.0, and 2.0 times Earth's gravity under the same air mass<br>• Different values of background air ($N_2$) pressure: 0.5, 1.0, and 5 bar under Earth's gravity | 230 to 500 K | This study |